\documentclass[letterpaper,compsoc,twoside]{IEEEtran}
\usepackage{fixltx2e} 
\usepackage{cmap} 
\usepackage{ifthen}
\usepackage[T1]{fontenc}
\usepackage[utf8]{inputenc}
\usepackage{amsmath}

\usepackage[font={small,it},labelfont=bf]{caption}
\usepackage{float}

\usepackage{longtable,ltcaption,array}
\newlength{\DUtablewidth} 

\pdfoutput=1
\usepackage{scipy}
\makeatletter
\def\PY@reset{\let\PY@it=\relax \let\PY@bf=\relax%
    \let\PY@ul=\relax \let\PY@tc=\relax%
    \let\PY@bc=\relax \let\PY@ff=\relax}
\def\PY@tok#1{\csname PY@tok@#1\endcsname}
\def\PY@toks#1+{\ifx\relax#1\empty\else%
    \PY@tok{#1}\expandafter\PY@toks\fi}
\def\PY@do#1{\PY@bc{\PY@tc{\PY@ul{%
    \PY@it{\PY@bf{\PY@ff{#1}}}}}}}
\def\PY#1#2{\PY@reset\PY@toks#1+\relax+\PY@do{#2}}

\expandafter\def\csname PY@tok@gd\endcsname{\def\PY@tc##1{\textcolor[rgb]{0.63,0.00,0.00}{##1}}}
\expandafter\def\csname PY@tok@gu\endcsname{\let\PY@bf=\textbf\def\PY@tc##1{\textcolor[rgb]{0.50,0.00,0.50}{##1}}}
\expandafter\def\csname PY@tok@gt\endcsname{\def\PY@tc##1{\textcolor[rgb]{0.00,0.27,0.87}{##1}}}
\expandafter\def\csname PY@tok@gs\endcsname{\let\PY@bf=\textbf}
\expandafter\def\csname PY@tok@gr\endcsname{\def\PY@tc##1{\textcolor[rgb]{1.00,0.00,0.00}{##1}}}
\expandafter\def\csname PY@tok@cm\endcsname{\let\PY@it=\textit\def\PY@tc##1{\textcolor[rgb]{0.25,0.50,0.56}{##1}}}
\expandafter\def\csname PY@tok@vg\endcsname{\def\PY@tc##1{\textcolor[rgb]{0.73,0.38,0.84}{##1}}}
\expandafter\def\csname PY@tok@m\endcsname{\def\PY@tc##1{\textcolor[rgb]{0.13,0.50,0.31}{##1}}}
\expandafter\def\csname PY@tok@mh\endcsname{\def\PY@tc##1{\textcolor[rgb]{0.13,0.50,0.31}{##1}}}
\expandafter\def\csname PY@tok@cs\endcsname{\def\PY@tc##1{\textcolor[rgb]{0.25,0.50,0.56}{##1}}\def\PY@bc##1{\setlength{\fboxsep}{0pt}\colorbox[rgb]{1.00,0.94,0.94}{\strut ##1}}}
\expandafter\def\csname PY@tok@ge\endcsname{\let\PY@it=\textit}
\expandafter\def\csname PY@tok@vc\endcsname{\def\PY@tc##1{\textcolor[rgb]{0.73,0.38,0.84}{##1}}}
\expandafter\def\csname PY@tok@il\endcsname{\def\PY@tc##1{\textcolor[rgb]{0.13,0.50,0.31}{##1}}}
\expandafter\def\csname PY@tok@go\endcsname{\def\PY@tc##1{\textcolor[rgb]{0.20,0.20,0.20}{##1}}}
\expandafter\def\csname PY@tok@cp\endcsname{\def\PY@tc##1{\textcolor[rgb]{0.00,0.44,0.13}{##1}}}
\expandafter\def\csname PY@tok@gi\endcsname{\def\PY@tc##1{\textcolor[rgb]{0.00,0.63,0.00}{##1}}}
\expandafter\def\csname PY@tok@gh\endcsname{\let\PY@bf=\textbf\def\PY@tc##1{\textcolor[rgb]{0.00,0.00,0.50}{##1}}}
\expandafter\def\csname PY@tok@ni\endcsname{\let\PY@bf=\textbf\def\PY@tc##1{\textcolor[rgb]{0.84,0.33,0.22}{##1}}}
\expandafter\def\csname PY@tok@nl\endcsname{\let\PY@bf=\textbf\def\PY@tc##1{\textcolor[rgb]{0.00,0.13,0.44}{##1}}}
\expandafter\def\csname PY@tok@nn\endcsname{\let\PY@bf=\textbf\def\PY@tc##1{\textcolor[rgb]{0.05,0.52,0.71}{##1}}}
\expandafter\def\csname PY@tok@no\endcsname{\def\PY@tc##1{\textcolor[rgb]{0.38,0.68,0.84}{##1}}}
\expandafter\def\csname PY@tok@na\endcsname{\def\PY@tc##1{\textcolor[rgb]{0.25,0.44,0.63}{##1}}}
\expandafter\def\csname PY@tok@nb\endcsname{\def\PY@tc##1{\textcolor[rgb]{0.00,0.44,0.13}{##1}}}
\expandafter\def\csname PY@tok@nc\endcsname{\let\PY@bf=\textbf\def\PY@tc##1{\textcolor[rgb]{0.05,0.52,0.71}{##1}}}
\expandafter\def\csname PY@tok@nd\endcsname{\let\PY@bf=\textbf\def\PY@tc##1{\textcolor[rgb]{0.33,0.33,0.33}{##1}}}
\expandafter\def\csname PY@tok@ne\endcsname{\def\PY@tc##1{\textcolor[rgb]{0.00,0.44,0.13}{##1}}}
\expandafter\def\csname PY@tok@nf\endcsname{\def\PY@tc##1{\textcolor[rgb]{0.02,0.16,0.49}{##1}}}
\expandafter\def\csname PY@tok@si\endcsname{\let\PY@it=\textit\def\PY@tc##1{\textcolor[rgb]{0.44,0.63,0.82}{##1}}}
\expandafter\def\csname PY@tok@s2\endcsname{\def\PY@tc##1{\textcolor[rgb]{0.25,0.44,0.63}{##1}}}
\expandafter\def\csname PY@tok@vi\endcsname{\def\PY@tc##1{\textcolor[rgb]{0.73,0.38,0.84}{##1}}}
\expandafter\def\csname PY@tok@nt\endcsname{\let\PY@bf=\textbf\def\PY@tc##1{\textcolor[rgb]{0.02,0.16,0.45}{##1}}}
\expandafter\def\csname PY@tok@nv\endcsname{\def\PY@tc##1{\textcolor[rgb]{0.73,0.38,0.84}{##1}}}
\expandafter\def\csname PY@tok@s1\endcsname{\def\PY@tc##1{\textcolor[rgb]{0.25,0.44,0.63}{##1}}}
\expandafter\def\csname PY@tok@gp\endcsname{\let\PY@bf=\textbf\def\PY@tc##1{\textcolor[rgb]{0.78,0.36,0.04}{##1}}}
\expandafter\def\csname PY@tok@sh\endcsname{\def\PY@tc##1{\textcolor[rgb]{0.25,0.44,0.63}{##1}}}
\expandafter\def\csname PY@tok@ow\endcsname{\let\PY@bf=\textbf\def\PY@tc##1{\textcolor[rgb]{0.00,0.44,0.13}{##1}}}
\expandafter\def\csname PY@tok@sx\endcsname{\def\PY@tc##1{\textcolor[rgb]{0.78,0.36,0.04}{##1}}}
\expandafter\def\csname PY@tok@bp\endcsname{\def\PY@tc##1{\textcolor[rgb]{0.00,0.44,0.13}{##1}}}
\expandafter\def\csname PY@tok@c1\endcsname{\let\PY@it=\textit\def\PY@tc##1{\textcolor[rgb]{0.25,0.50,0.56}{##1}}}
\expandafter\def\csname PY@tok@kc\endcsname{\let\PY@bf=\textbf\def\PY@tc##1{\textcolor[rgb]{0.00,0.44,0.13}{##1}}}
\expandafter\def\csname PY@tok@c\endcsname{\let\PY@it=\textit\def\PY@tc##1{\textcolor[rgb]{0.25,0.50,0.56}{##1}}}
\expandafter\def\csname PY@tok@mf\endcsname{\def\PY@tc##1{\textcolor[rgb]{0.13,0.50,0.31}{##1}}}
\expandafter\def\csname PY@tok@err\endcsname{\def\PY@bc##1{\setlength{\fboxsep}{0pt}\fcolorbox[rgb]{1.00,0.00,0.00}{1,1,1}{\strut ##1}}}
\expandafter\def\csname PY@tok@kd\endcsname{\let\PY@bf=\textbf\def\PY@tc##1{\textcolor[rgb]{0.00,0.44,0.13}{##1}}}
\expandafter\def\csname PY@tok@ss\endcsname{\def\PY@tc##1{\textcolor[rgb]{0.32,0.47,0.09}{##1}}}
\expandafter\def\csname PY@tok@sr\endcsname{\def\PY@tc##1{\textcolor[rgb]{0.14,0.33,0.53}{##1}}}
\expandafter\def\csname PY@tok@mo\endcsname{\def\PY@tc##1{\textcolor[rgb]{0.13,0.50,0.31}{##1}}}
\expandafter\def\csname PY@tok@mi\endcsname{\def\PY@tc##1{\textcolor[rgb]{0.13,0.50,0.31}{##1}}}
\expandafter\def\csname PY@tok@kn\endcsname{\let\PY@bf=\textbf\def\PY@tc##1{\textcolor[rgb]{0.00,0.44,0.13}{##1}}}
\expandafter\def\csname PY@tok@o\endcsname{\def\PY@tc##1{\textcolor[rgb]{0.40,0.40,0.40}{##1}}}
\expandafter\def\csname PY@tok@kr\endcsname{\let\PY@bf=\textbf\def\PY@tc##1{\textcolor[rgb]{0.00,0.44,0.13}{##1}}}
\expandafter\def\csname PY@tok@s\endcsname{\def\PY@tc##1{\textcolor[rgb]{0.25,0.44,0.63}{##1}}}
\expandafter\def\csname PY@tok@kp\endcsname{\def\PY@tc##1{\textcolor[rgb]{0.00,0.44,0.13}{##1}}}
\expandafter\def\csname PY@tok@w\endcsname{\def\PY@tc##1{\textcolor[rgb]{0.73,0.73,0.73}{##1}}}
\expandafter\def\csname PY@tok@kt\endcsname{\def\PY@tc##1{\textcolor[rgb]{0.56,0.13,0.00}{##1}}}
\expandafter\def\csname PY@tok@sc\endcsname{\def\PY@tc##1{\textcolor[rgb]{0.25,0.44,0.63}{##1}}}
\expandafter\def\csname PY@tok@sb\endcsname{\def\PY@tc##1{\textcolor[rgb]{0.25,0.44,0.63}{##1}}}
\expandafter\def\csname PY@tok@k\endcsname{\let\PY@bf=\textbf\def\PY@tc##1{\textcolor[rgb]{0.00,0.44,0.13}{##1}}}
\expandafter\def\csname PY@tok@se\endcsname{\let\PY@bf=\textbf\def\PY@tc##1{\textcolor[rgb]{0.25,0.44,0.63}{##1}}}
\expandafter\def\csname PY@tok@sd\endcsname{\let\PY@it=\textit\def\PY@tc##1{\textcolor[rgb]{0.25,0.44,0.63}{##1}}}

\makeatother

\providecommand*{\DUfootnotemark}[3]{%
  \raisebox{1em}{\hypertarget{#1}{}}%
  \hyperlink{#2}{\textsuperscript{#3}}%
}
\providecommand{\DUfootnotetext}[4]{%
  \begingroup%
  \renewcommand{\thefootnote}{%
    \protect\raisebox{1em}{\protect\hypertarget{#1}{}}%
    \protect\hyperlink{#2}{#3}}%
  \footnotetext{#4}%
  \endgroup%
}

\providecommand*{\DUrole}[2]{%
  \ifcsname DUrole#1\endcsname%
    \csname DUrole#1\endcsname{#2}%
  \else
    \ifcsname docutilsrole#1\endcsname%
      \csname docutilsrole#1\endcsname{#2}%
    \else%
      #2%
    \fi%
  \fi%
}

\providecommand*{\DUroletitlereference}[1]{\textsl{#1}}

\ifthenelse{\isundefined{\hypersetup}}{
  \usepackage[colorlinks=true,linkcolor=blue,urlcolor=blue]{hyperref}
  \urlstyle{same} 
}{}

\begin{document}
\newcounter{footnotecounter}\title{Performance of Python runtimes on a non-numeric scientific code}\author{Riccardo Murri$^{\setcounter{footnotecounter}{1}\fnsymbol{footnotecounter}\setcounter{footnotecounter}{2}\fnsymbol{footnotecounter}}$%
          \setcounter{footnotecounter}{1}\thanks{\fnsymbol{footnotecounter} %
          Corresponding author: \protect\href{mailto:riccardo.murri@gmail.com}{riccardo.murri@gmail.com}}\setcounter{footnotecounter}{2}\thanks{\fnsymbol{footnotecounter} University of Zurich, Grid Computing Competence Center}\thanks{%

          \noindent%
          Copyright\,\copyright\,2014 Riccardo Murri. This is an open-access article distributed under the terms of the Creative Commons Attribution License, which permits unrestricted use, distribution, and reproduction in any medium, provided the original author and source are credited. http://creativecommons.org/licenses/by/3.0/%
        }}\maketitle
          \renewcommand{\leftmark}{PROC. OF THE 6th EUR. CONF. ON PYTHON IN SCIENCE (EUROSCIPY 2013)}
          \renewcommand{\rightmark}{PERFORMANCE OF PYTHON RUNTIMES ON A NON-NUMERIC SCIENTIFIC CODE}

\setcounter{page}{45}
\newcommand*{\docutilsroleref}{\ref}
\newcommand*{\docutilsrolelabel}{\label}
\AtEndDocument{\cleardoublepage}
\begin{abstract}The Python library FatGHol \cite{FatGHoL} used in \cite{Murri2012} to reckon
the rational homology of the moduli space of Riemann surfaces is an
example of a non-numeric scientific code: most of the processing it
does is generating graphs (represented by complex Python objects)
and computing their isomorphisms (a triple of Python lists; again a
nested data structure). These operations are repeated many times
over: for example, the spaces $M_{0,6}$ and $M_{1,4}$
are triangulated by 4'583'322 and 747'664 graphs, respectively.

This is an opportunity for every Python runtime to prove its
strength in optimization. The purpose of this experiment was to
assess the maturity of alternative Python runtimes, in terms of:
compatibility with the language as implemented in CPython 2.7, and
performance speedup.

This paper compares the results and experiences from running
FatGHol with different Python runtimes: CPython 2.7.5, PyPy 2.1,
Cython 0.19, Numba 0.11, Nuitka 0.4.4 and Falcon.\end{abstract}

\section{Introduction%
  \label{introduction}%
}

The moduli space $M_{g,n}$ of smooth Riemann surfaces is a
topological space which has been subject of much research both in
algebraic geometry and in string theory. It is known since the '90s
that this space has a triangulation indexed by a special kind of
graphs \cite{Penner1988,Kontsevich1992,ConantVogtmann2003},
nicknamed \textquotedbl{}fat graphs\textquotedbl{}.

Since graphs are combinatorial and discrete objects, a computational
approach to the problem of computing topological invariants of
$M_{g,n}$ is now feasible; algorithms to enumerate fatgraphs and
compute their graph homology have been devised in \cite{Murri2012} and
implemented in Python.

We propose an experiment whose purpose is to assess the maturity of
alternative Python runtimes, in terms of:\newcounter{listcnt0}
\begin{list}{(\alph{listcnt0})}
{
\usecounter{listcnt0}
\setlength{\rightmargin}{\leftmargin}
}

\item 

compatibility with the language as implemented in CPython 2.7, and
\item 

performance speedup.\end{list}

In particular, we were interested in the possible speedups of
a large non-numeric code.

\section{Experiment setup%
  \label{experiment-setup}%
}

The \DUroletitlereference{FatGHoL <http://fatghol.googlecode.com/>} \cite{FatGHoL} program was
used as a test code.  FatGHoL computes homology of the moduli spaces
of Riemann surfaces $M_{g,n}$ via Penner-Kontsevich's fatgraph
simplicial complex \cite{Penner1988,Kontsevich1992}.  Homology is one
of the most important invariants of topological spaces.  There are
several homology theories but they all share this computational
procedure outline: given a vector space of (generalized) \emph{simplex
chains} and a \emph{boundary operator}, which is by definition a linear
operator $D$ such that $D^2=0$, the homology space is by
definition $\mathop{\textrm{Ker }} D / \mathop{\textrm{Im }} D$.  In graph homology, however, it is the
computation of these simplices and boundary that takes up the largest
fraction of compute time: the simplex chains are defined as formal
linear combinations of graphs, and the boundary operator maps a graph
into a linear combination of graphs obtained by contracting its edges.
Thus, explicit construction of the simplices requires enumerating all
distinct isomorphism classes of fatgraphs, and then computing their
mutual relationships upon contraction of edges.

The FatGHoL program runs in three stages:\setcounter{listcnt0}{0}
\begin{list}{\arabic{listcnt0}.}
{
\usecounter{listcnt0}
\setlength{\rightmargin}{\leftmargin}
}

\item 

generate fatgraphs,
\item 

explicitly compute the boundary operator in matrix form,
\item 

actually solve the homology linear system.\end{list}

The last step has been disabled in the test code as it is implemented
in C++ for speed.  What remains is 100\% pure Python code that runs on
Python 2.6+ (but could run on 2.5 with minimal modifications).

FatGHoL involves a large number of graph isomorphism computations:
especially during fatgraph generation, each candidate fatgraph needs
to be compared to all fatgraphs already discovered, in order to avoid
duplicates. In later stages, the isomorphism computations are cached
in memory, but in step 2 additional data is created for each graph,
in order to pass from fatgraphs to simplices.

It is worth noting that the FatGHoL code exercises many of Python's
advanced data manipulation features, like list and dictionary
comprehensions, slicing, etc. but does not use any kind of tight
nested loops of the kind normally featured in numeric codes.

Profiling data show more precisely how much work is done at the Python
level in the simplest case $M_{0,4}$.  The following listing
shows profiling data extracted from a complete run of FatGHoL on
CPython 2.7.5; 15787953 function calls (15728052 primitive calls) were
effected in 39.572 seconds; the top 10 most called functions, ordered
by call count are:%
\begin{quote}\begin{verbatim}
   ncalls   tottime  filename:lineno(function)
  2216088     2.175  rg.py:227(<genexpr>)
   966575     0.819  rg.py:143(is_loop)
   775362     0.839  cyclicseq.py:88(__getitem__)
   775362     0.634  rg.py:170(other_end)
   722308     3.438  combinatorics.py:368(__init__)
   539039     1.689  cyclicseq.py:112(__getslice__)
506075/447917 0.745  cache.py:181(wrapper)
   476134     1.122  combinatorics.py:441(rearranged)
   385725     0.355  rg.py:137(__init__)
   345740     0.849  rg.py:568(_first_unused_corner)
\end{verbatim}

\end{quote}
The FatGHoL code was run with seven different alternative Python
runtimes:%
\begin{itemize}

\item 

CPython 2.7.5;
\item 

Cython 0.19.1;
\item 

Cython 0.19.1 in \textquotedbl{}pure Python mode\textquotedbl{};
\item 

Falcon 0.05;
\item 

Nuitka 0.4.4;
\item 

PyPy 2.1;
\item 

Numba 0.10.0 and 0.11.0 with \texttt{@autojit}.
\end{itemize}

A detailed description of each of these is given in a later
section; Table \DUrole{ref}{tab-effort} provides an overview of the
installation and usage features of the different runtimes.
The code used to install the software and run the experiments is
available on GitHub at
\url{https://github.com/riccardomurri/python-runtimes-shootout}.\begin{table*}
\setlength{\DUtablewidth}{0.8\linewidth}
\begin{longtable*}[c]{|p{0.160\DUtablewidth}|p{0.160\DUtablewidth}|p{0.160\DUtablewidth}|p{0.160\DUtablewidth}|p{0.160\DUtablewidth}|p{0.160\DUtablewidth}|}
\hline

Runtime & 

\emph{Cython 0.19.1} & 

\emph{Falcon 0.05} & 

\emph{Nuitka 0.4.4} & 

\emph{Numba 0.11.0} & 

\emph{PyPy 2.1} \\
\hline

\emph{Installed size} (MB) & 

30 $^a$ & 

14 $^a$ & 

25 $^a$ & 

97 $^a$ (+ 518MB of LLVM 3.2) & 

162 $^b$ \\
\hline

\emph{Install script length} (SLOC) & 

6 & 

8 & 

10 & 

24 & 

19 \\
\hline

\emph{Usage documentation} & 

extensive & 

minimal & 

short how-to to explain the different compilation options available & 

minimal, mostly examples & 

none \\
\hline

\emph{Porting/optimization documentation} & 

extensive & 

none & 

list of optimizations that the runtime does (or will) support & 

none & 

provides only a list of compatibility issues; I could find no
list of \emph{Do}-s and \emph{Don't}-s for better speed in PyPy \\
\hline

\emph{Porting/optimization effort} & 

none (\textquotedbl{}Pure Python\textquotedbl{} mode) to very heavy (\texttt{.pxd} hinting) & 

none: runs unmodified Python code & 

none: runs unmodified Python code & 

light (w/ \texttt{@autojit}) to heavy (\texttt{@jit} with types) & 

none: runs unmodified Python code \\
\hline
\end{longtable*}
\caption{\DUrole{label}{tab-effort} Comparison of installation features of the
             Python runtimes. $^a$ Plus 123MB for the CPython
             interpreter, which is anyway needed. $^b$ Does not need
             the CPython interpreter in addition, as all others do.}\end{table*}

Except for Cython in \textquotedbl{}pure Python mode\textquotedbl{} and Numba, all runtimes run
the unmodified Python code of FatGHoL.  Cython in \textquotedbl{}pure Python mode\textquotedbl{}
requires the addition of decorators to the Python code that specify
the types of function arguments and local variables to increase
speedup of selected portions of the code.  Similarly, Numba uses the
decorators \texttt{@jit} or \texttt{@autojit} to mark functions that should be
compiled to native code (the \href{http://nbviewer.ipython.org/gist/Juanlu001/3914904}{difference between the two decorators}
is that that \texttt{@autojit} infers types at runtime, whereas \texttt{@jit}
requires the programmer to specify them\DUfootnotemark{id10}{no-more-autojit}{1}); we only
used the \texttt{@autojit} decorator to mark the same functions that were
marked as optimization candidates in the Cython experiment.%
\DUfootnotetext{no-more-autojit}{id10}{1}{\phantomsection\label{no-more-autojit}
Note that in more recent versions of Numba, the
two decorators have been fused into one:
\texttt{@jit} uses the supplied types, or infers them
if none are given.}

Each Python runtime was run on 4 test cases: computing the homology of
the $M_{0,4}$, $M_{0,5}$, $M_{1,3}$, and
$M_{2,1}$ moduli spaces.  The test cases take from 0.20s to more
than 2 minutes of runtime with CPython 2.7.  Each test case was run 10
times and the best time and lowest RAM occupation are reported in the
summary tables below.

\section{Results%
  \label{results}%
}

Falcon and Numba could not run the code (see details in a later
section) and thus do not appear in the report below.

For each runtime, the total used CPU time and memory were measured:
results and summary graphs are given in figures \DUrole{ref}{cpu-all} and
\DUrole{ref}{mem-all}.  Detailed comparisons are given in the other figures.\begin{figure*}[tb]\noindent\makebox[\textwidth][c]{\includegraphics[scale=0.45]{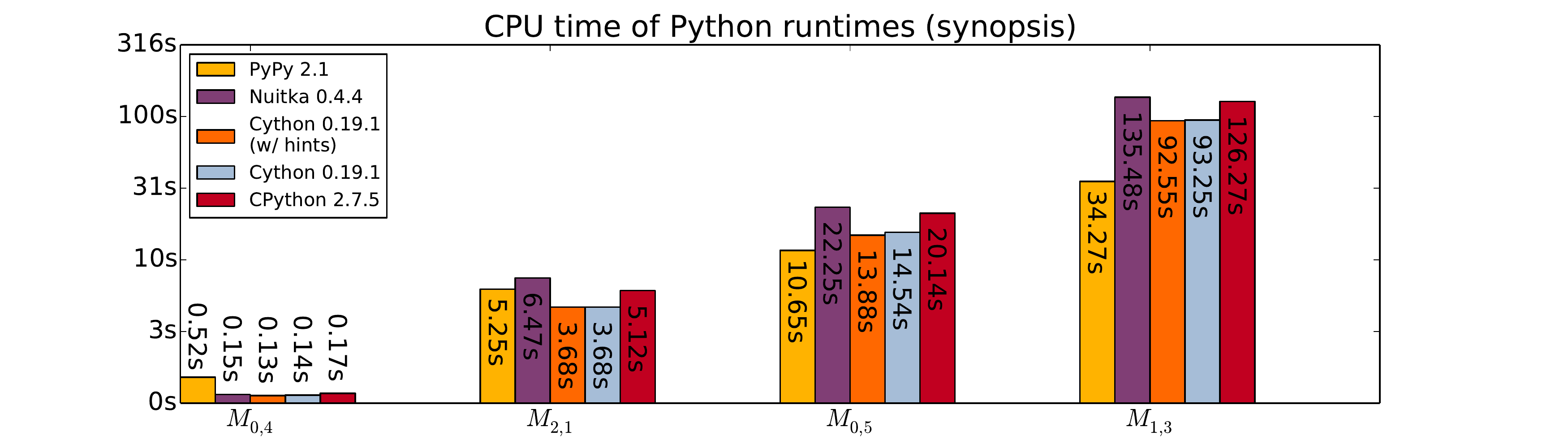}}
\caption{Comparison of the total CPU time used by each runtime on the
different test cases.  The $x$-axis is sorted so that the
runtimes for CPython 2.7.5 are ascending.  The $y$-axis shows
values in seconds (smaller is better). Note that the $y$-axis
is drawn on a logarithmic scale!
\DUrole{label}{cpu-all}}
\end{figure*}

The CPU time data prompt a few observations:%
\begin{itemize}

\item 

PyPy gives the best results, provided the code runs long enough to
discount for the startup time of the JIT compiler. Given
enough time, the JIT compiler gives extremely good results, with
speedups of 100\% to 400\% relative to CPython in the $M_{0,5}$
and $M_{1,3}$ cases.  In other words, for the JIT approach to
pay off, the code needs to perform many iterations of the same code
path (this is certainly the case for FatGHoL), because compiling a
single function to native code takes a non-negligible amount of
time.  The break-even point for the FatGHoL code seems to be around
5 seconds of runtime: on $M_{2,1}$, the CPU time taken by
CPython and PyPy are almost equal.
\item 

Cython gives consistently about a 30\% speedup on unmodified Python
code.  However, the \textquotedbl{}pure Python mode\textquotedbl{}, in which Cython takes
variable typing hints embedded in the code does not seem to give any
advantage: results of the two runs are not significantly different.
This might be related to a bug in the current version of Cython, see
details in a later section.
\end{itemize}
\begin{figure*}[tb]\noindent\makebox[\textwidth][c]{\includegraphics[scale=0.45]{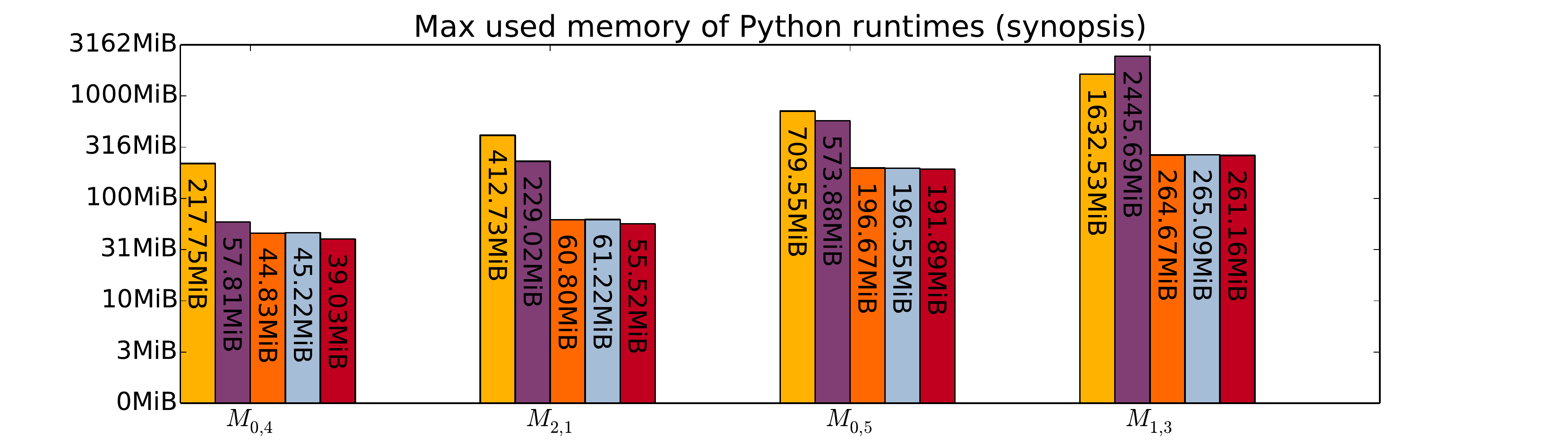}}
\caption{Comparison of the total RAM used by each runtime on the
different test cases.  The $x$-axis is sorted so that the
RAM usage for CPython 2.7.5 are ascending.  The $y$-axis
shows values in MBs (smaller is better).  Note that the $y$-axis
is drawn on a logarithmic scale!
\DUrole{label}{mem-all}}
\end{figure*}

The large memory consumption from PyPy and Nuitka stands out in the
memory data of figure \DUrole{ref}{mem-all}.  On the other hand, there is no
significant increase in memory usage between CPython and Cython.

The large memory usage of PyPy can be explained by the fact that the
JIT infrastructure must keep in memory the profile and traces for all
the code paths taken.  In any long-running program, the memory should
eventually reach a steady state and not increase any further; it
should be noted however, that in these benchmarks the memory used by
the PyPy JIT framework dwarfs the memory used by the program itself.

We have no explanation for the large memory consumption of Nuitka.\begin{figure}[tbp]\noindent\makebox[\columnwidth][l]{\includegraphics[scale=0.40]{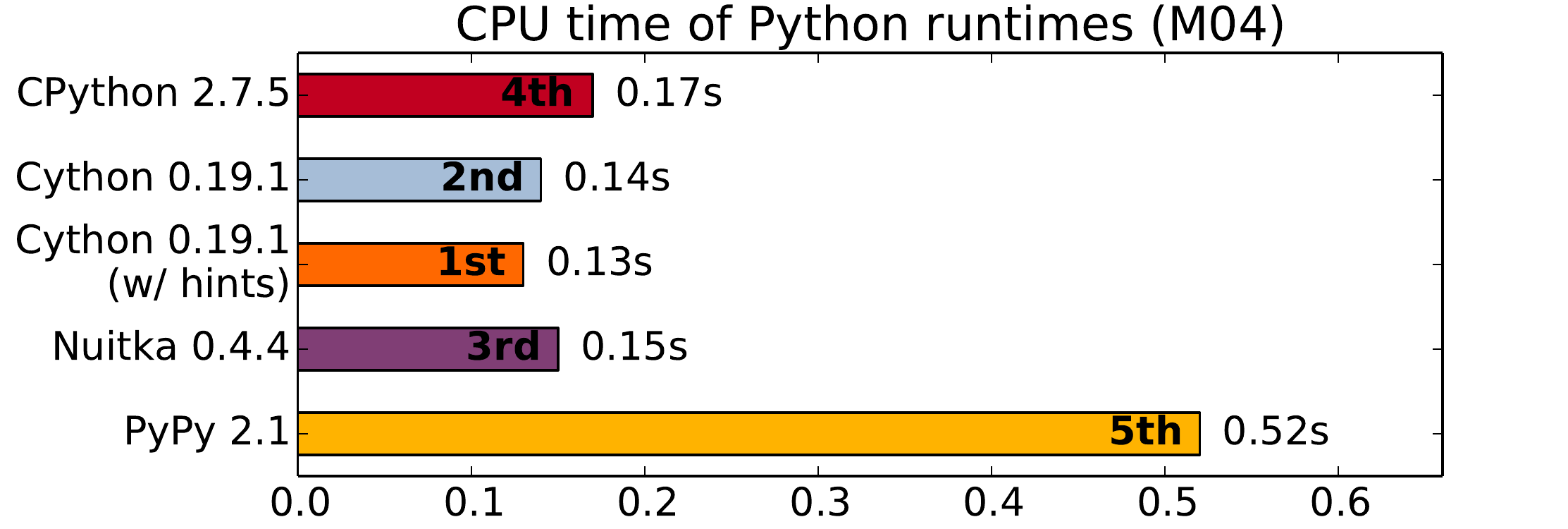}}
\caption{Comparison of the total CPU time used by each runtime on the
$M_{0,4}$ test case.  The $x$-axis shows
values in seconds.
\DUrole{label}{cpu-M04}}
\end{figure}\begin{figure}[tbp]\noindent\makebox[\columnwidth][l]{\includegraphics[scale=0.40]{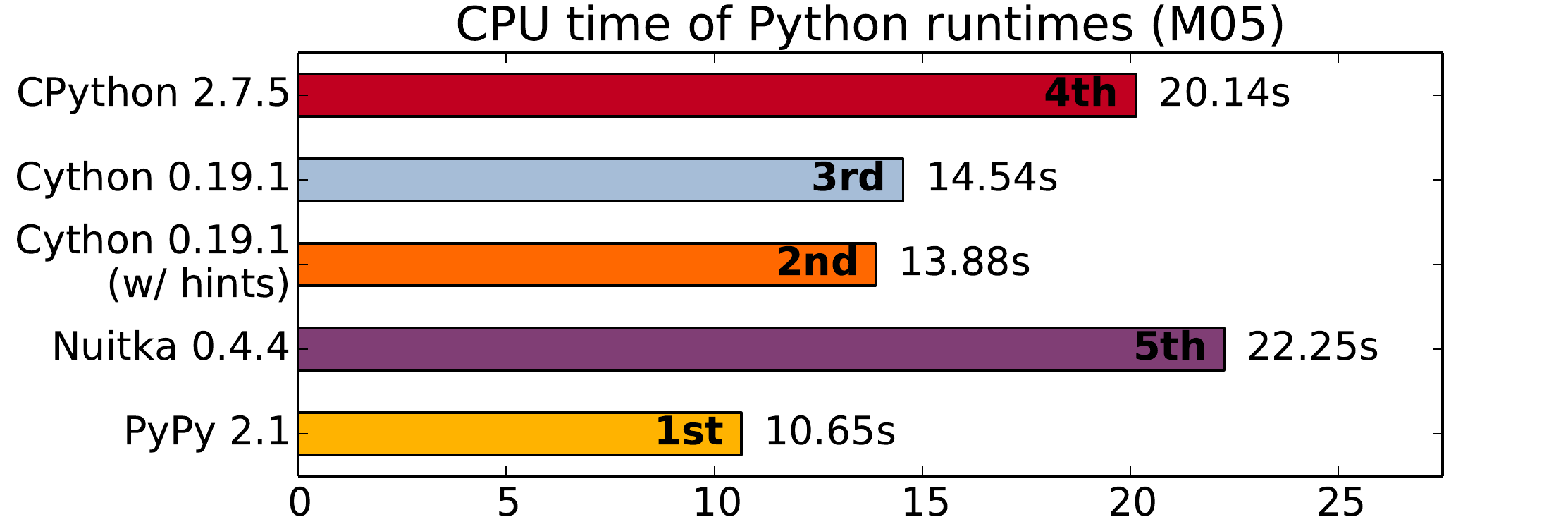}}
\caption{Comparison of the total CPU time used by each runtime on the
$M_{0,5}$ test case.  The $x$-axis shows
values in seconds.
\DUrole{label}{cpu-M05}}
\end{figure}\begin{figure}[tbp]\noindent\makebox[\columnwidth][l]{\includegraphics[scale=0.40]{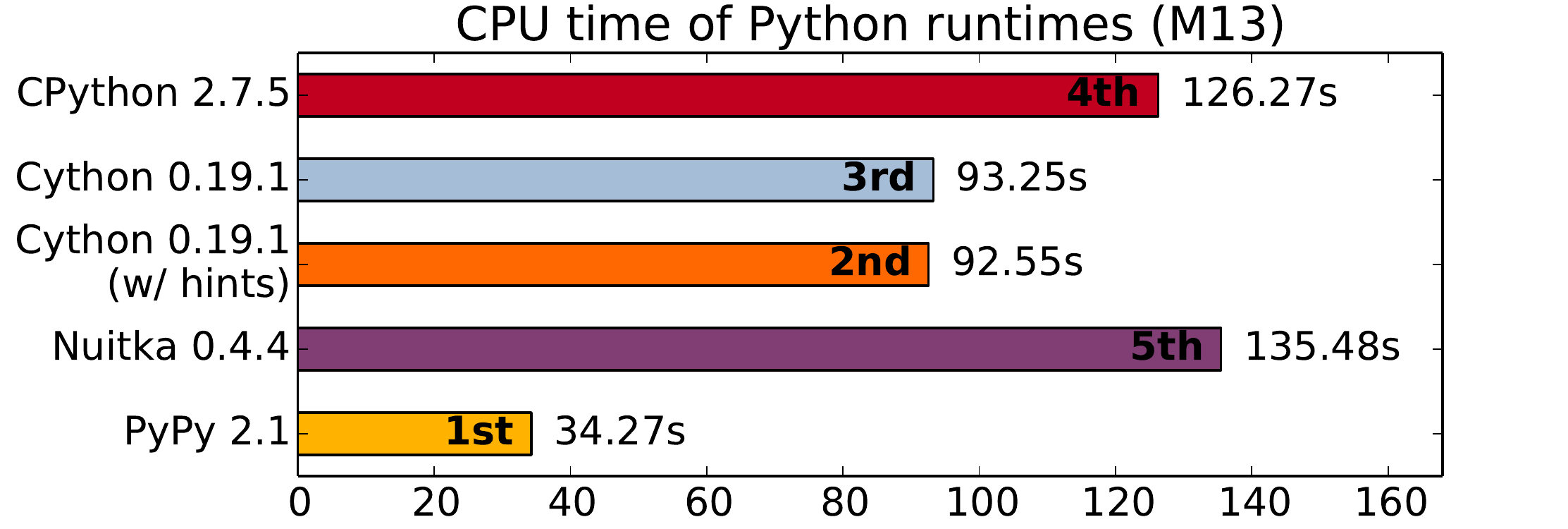}}
\caption{Comparison of the total CPU time used by each runtime on the
$M_{1,3}$ test case.  The $x$-axis shows
values in seconds.
\DUrole{label}{cpu-M13}}
\end{figure}\begin{figure}[tbp]\noindent\makebox[\columnwidth][l]{\includegraphics[scale=0.40]{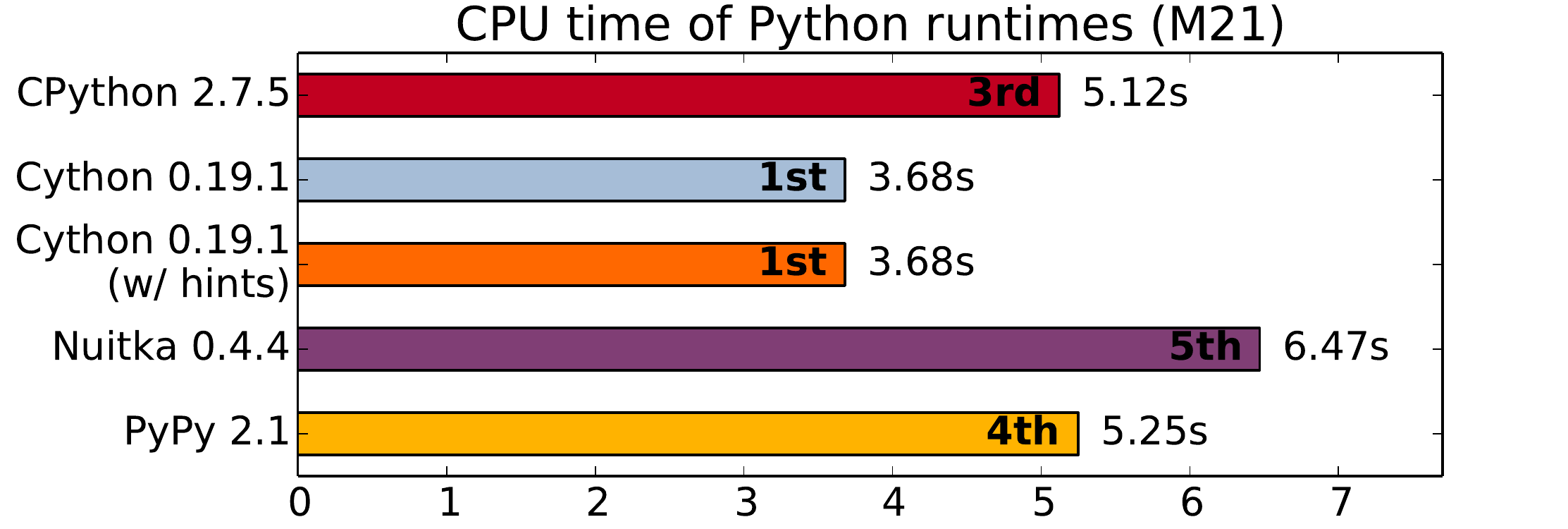}}
\caption{Comparison of the total CPU time used by each runtime on the
$M_{2,1}$ test case.  The $x$-axis shows
values in seconds.
\DUrole{label}{cpu-M21}}
\end{figure}\begin{figure}[tbp]\noindent\makebox[\columnwidth][l]{\includegraphics[scale=0.40]{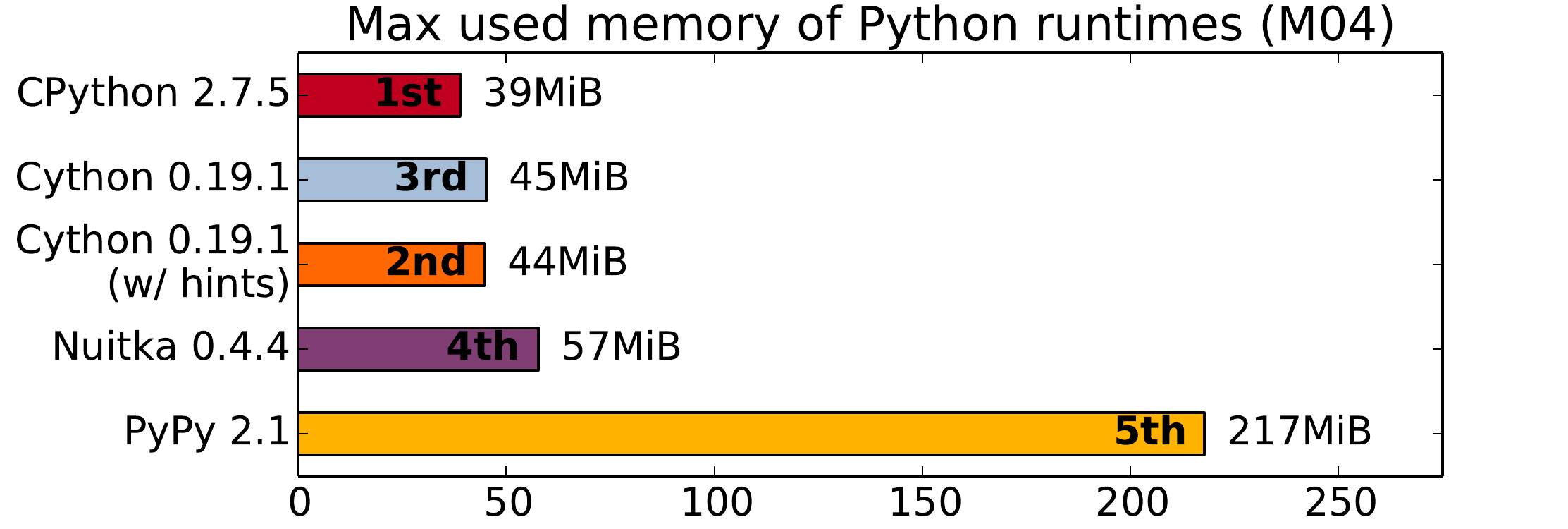}}
\caption{Comparison of the total RAM usage by each runtime on the
$M_{0,4}$ test case.  The $x$-axis shows
values in MBs.
\DUrole{label}{mem-M04}}
\end{figure}\begin{figure}[tbp]\noindent\makebox[\columnwidth][l]{\includegraphics[scale=0.40]{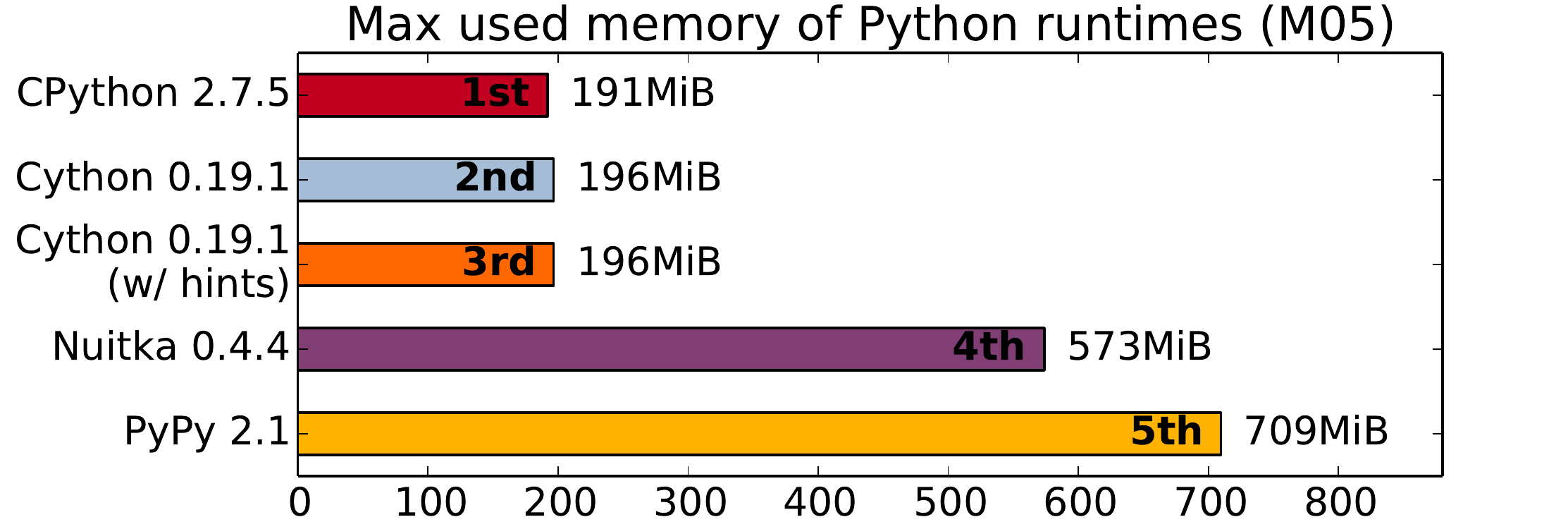}}
\caption{Comparison of the total RAM usage by each runtime on the
$M_{0,5}$ test case.  The $x$-axis shows
values in MBs.
\DUrole{label}{mem-M05}}
\end{figure}\begin{figure}[tbp]\noindent\makebox[\columnwidth][l]{\includegraphics[scale=0.40]{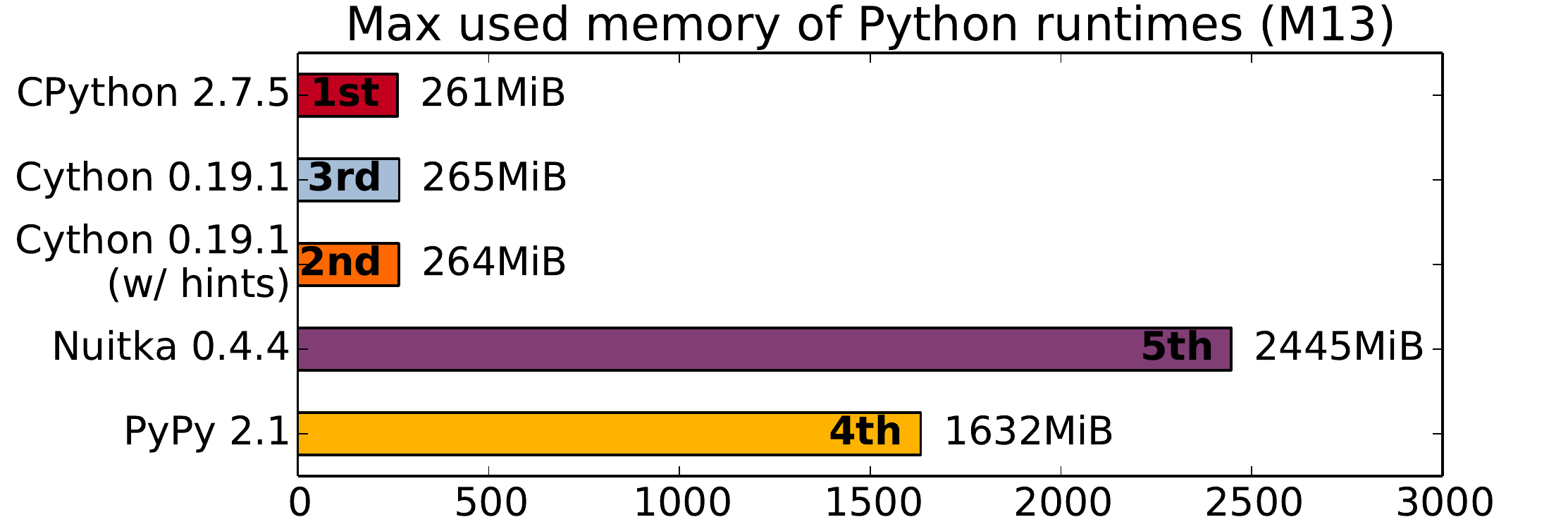}}
\caption{Comparison of the total RAM usage by each runtime on the
$M_{1,3}$ test case.  The $x$-axis shows
values in MBs.
\DUrole{label}{mem-M13}}
\end{figure}\begin{figure}[tbp]\noindent\makebox[\columnwidth][l]{\includegraphics[scale=0.40]{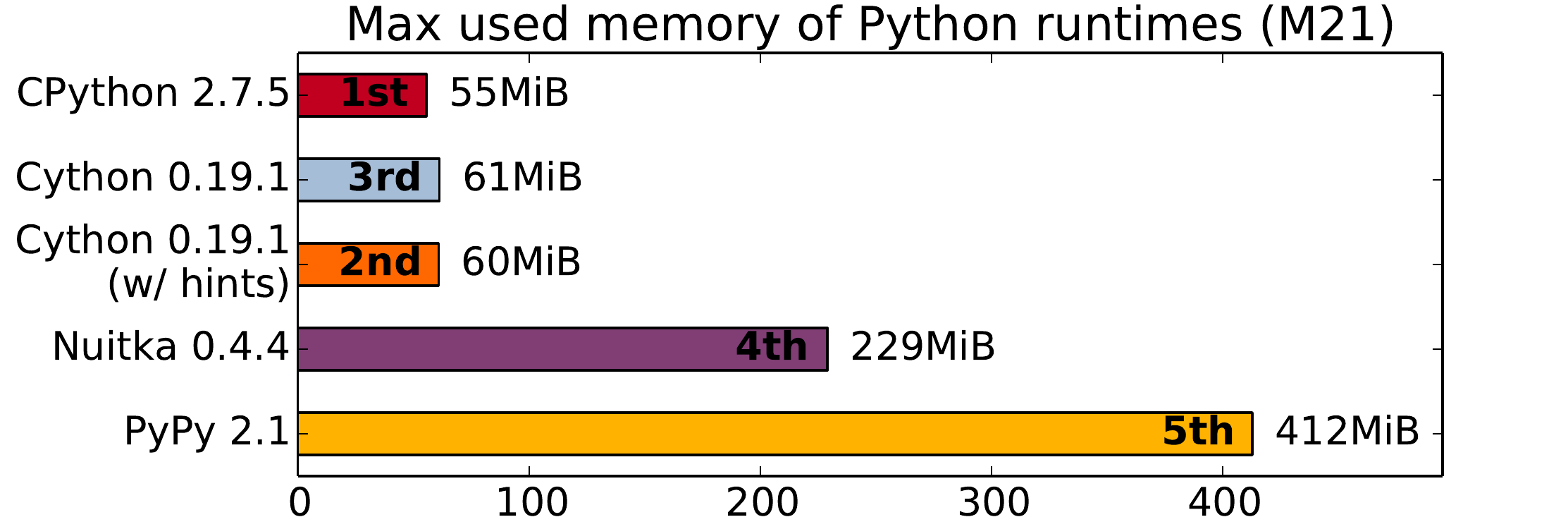}}
\caption{Comparison of the total RAM usage by each runtime on the
$M_{2,1}$ test case.  The $x$-axis shows
values in MBs.
\DUrole{label}{mem-M21}}
\end{figure}

\section{Runtime systems details%
  \label{runtime-systems-details}%
}

\subsection{\DUroletitlereference{Cython 0.19.1 <http://cython.org/>}%
  \label{cython-0-19-1-http-cython-org}%
}

Cython is a compiler for a superset of the Python language. It
translates Python modules to a C or C++ source that is then compiled
to a native code library that CPython can load and use. Cython
optimizes best when users decorate the source code with hints at the
types of variables and functions; it can also translate unmodified
Python code, but then no type inference is performed. Cython allows a
variety of ways for giving these type hints; its so-called \textquotedbl{}pure
Python\textquotedbl{} mode requires users to insert functions and variable
decorators in the code: the Cython compiler can act on these
directives, but the CPython interpreter will instead load a \texttt{cython}
module which turns them into no-ops.

We tested Cython twice: on the unmodified Python sources, and with
hinting in the \textquotedbl{}pure Python\textquotedbl{} mode.  The graphs show however very
little difference between the two modes; this could be a consequence
of Cython \href{http://trac.cython.org/cython_trac/ticket/477}{defect ticket \#477}.

Cython does its best when the source code is annotated with its
extended keywords, which allow specifying the types of variables
(which allows optimizations, e.g., in loops), or
marking certain functions as C-only (which saves time when
dereferencing variables).  This extended markup can be provided either
in the sources, or in additional \texttt{.pxd} files.  We have not done
this exercise, however, as the amount of coding time required to
properly mark all functions and variables is quite substantial.

\subsection{\DUroletitlereference{Falcon 0.05 <https://github.com/rjpower/falcon>}%
  \label{falcon-0-05-https-github-com-rjpower-falcon}%
}

Falcon is a Python extension module that hacks into a CPython
interpreter and changes the execution loop, implementing several
optimizations (for instance, using a register-based VM instead of a
stack-based one) that the Falcon authors think should be used in the
upstream CPython interpreter too. However, Falcon is still in early
stages of development and crashes on FatGHoL code with a segmentation
fault.

\subsection{\DUroletitlereference{Numba <http://numba.pydata.org/>}%
  \label{numba-http-numba-pydata-org}%
}

As its website states:\begin{quotation}%
\begin{quote}

Numba is an optimizing compiler for Python; it uses the LLVM
compiler infrastructure to compile Python syntax to machine code.
It is NumPy-aware and can speed up code using NumPy arrays.  Other,
less well-typed code will be translated to Python C-API calls
effectively removing the \textquotedbl{}interpreter\textquotedbl{} but not removing the dynamic
indirection. Numba is also not a Just-In-Time compiler.
\end{quote}
\end{quotation}

Numba requires the code developer to use either the \texttt{@autojit} (use
run-time type info) or the \texttt{@jit} (explicitly provide type
information) decorators to mark those functions that should be
compiled. For our experiment, we used the decorator \texttt{@autojit} on
all functions that were decorated also in the Cython test.

Versions 0.10.0 and 0.11.0 of Numba were tested; we could not get
either version to work.

Numba version 0.10.0 dies with an internal error (\textquotedbl{}TypeError: type\_container() takes exactly 1 argument (3 given)\textquotedbl{}, reported as
\href{https://github.com/numba/numba/issues/295}{Issue \#295} on Numba's GitHub issue tracker), that has
been fixed in version 0.11.

However, Numba 0.11.0 with a \textquotedbl{}NotImplementedError: Unable to cast from
\{ i64, i8* \}* to \{ i64, i8* \}\textquotedbl{} message.  This has been reported as
\href{https://github.com/numba/numba/issues/350}{Issue \#350} on the \href{https://github.com/numba/numba/issues?state=open}{issue tracker} and is waiting for a fix.

\subsection{\DUroletitlereference{Nuitka 0.4.4 <http://www.nuitka.net/>}%
  \label{nuitka-0-4-4-http-www-nuitka-net}%
}

Nuitka translates Python (2.6+) into a C++ program that then uses
\texttt{libpython} to execute in the same way as CPython does, in a very
compatible way.  Although still in development, Nuitka claims that it
already:\begin{quotation}%
\begin{quote}

create{[}s{]} the most efficient native code from this. This
means to be fast with the basic Python object handling.
\end{quote}
\end{quotation}

Results of this experiment do not seem to corroborate this claim.

\subsection{\DUroletitlereference{PyPy 2.1 <http://pypy.org/>}%
  \label{pypy-2-1-http-pypy-org}%
}

PyPy is a Python language interpreter with a Just-In-Time compiler
(and many other features!).  It can thus translate repetitive Python
code into native code on the fly.  PyPy must first be bootstrapped by
compiling itself, which takes a lot of time and RAM, but then it is a
drop-in replacement for the \texttt{python} command and just works.

\section{Acknowledgements%
  \label{acknowledgements}%
}

The author acknowledges support of the Informatik Dienste of the
University of Zurich, particularly for the usage of the new SGI UV 2000
machine for running the tests.  I would also like to thank Kay Hayen,
Marc Florisson, Russel Power and Alex Rubynstein for their readiness
to discuss and fix the bugs I reported on Nuitka, Numba, and Falcon.
Finally, I would like to express my gratitude to all those who made
remarks and inquiries at the EuroSciPy poster session, and
particularly Ronan Lamy and Denis Engemann for their insightful
comments.  Finally, I would like to thank Mike Mueller and Pierre
de Buyl for reviewing the initial draft paper and making many useful
suggestions for improving it.

\end{document}